\documentclass[floatfix,superscriptaddress,showpacs,amssymb,10pt,aps,prd,reprint,longbibliography]{revtex4-1}

\usepackage{graphicx,epsfig,amssymb} 
\usepackage{amsmath,amsfonts, times}
\usepackage{bm} 

\usepackage[linktocpage,colorlinks]{hyperref}
\usepackage[caption=false]{subfig}
\usepackage[usenames]{color}
\usepackage{natbib}
\usepackage{soul}
\usepackage[utf8x]{inputenc}
\usepackage{float}

\definecolor{coolblack}{rgb}{0.0, 0.18, 0.39}
\definecolor{darkred}{rgb}{0.5,0,0}
\definecolor{darkgreen}{rgb}{0,0.5,0}
\definecolor{darkblue}{rgb}{0,0,0.5}
\definecolor{lapislazuli}{rgb}{0.15, 0.38, 0.61}
\definecolor{venetianred}{rgb}{0.78, 0.03, 0.08}
\definecolor{bleudefrance}{rgb}{0.19, 0.55, 0.91}
\definecolor{dogwoodrose}{rgb}{0.84, 0.09, 0.41}
\hypersetup{colorlinks=true, citecolor=darkgreen, linkcolor=darkblue,
	urlcolor = blue}

\def\be{\begin{equation}}
	\def\ee{\end{equation}}

\newcommand{\bea}{\begin{eqnarray}}
	\newcommand{\eea}{\end{eqnarray}}
\newcommand{\ben}{\begin{enumerate}}
	\newcommand{\een}{\end{enumerate}}
\newcommand{\bi}{\begin{itemize}}
	\newcommand{\ei}{\end{itemize}}

\def\ga{\mathrel{\raise.3ex\hbox{$>$\kern-.75em\lower1ex\hbox{$\sim$}}}}
\def\la{\mathrel{\raise.3ex\hbox{$<$\kern-.75em\lower1ex\hbox{$\sim$}}}}

\def\be{\begin{equation}}
	\def\ee{\end{equation}}
\def\b{\begin{equation}}
	\def\e{\end{equation}}

\def\I_M{{I_{\scriptscriptstyle M\times M}}}

\def\be{\begin{equation}}
	\def\ee{\end{equation}}
\def\bea{\begin{eqnarray}}
	\def\eea{\end{eqnarray}}
\newcommand{\beq}{\begin{eqnarray}}
	\newcommand{\eeq}{\end{eqnarray}}

\newcommand{\beqal}{\begin{eqnarray}\label}
	\newcommand{\beqa}{\begin{eqnarray}}
		\newcommand{\eeqa}{\end{eqnarray}}

\begin{document}

\author{Qi-Quan Li}
\address{School of Physics Science and Technology, Xinjiang University, Urumqi 830046, China}

\author{Yu Zhang}
\email[Corresponding author:~]{zhangyu\_128@126.com}
\address{Faculty of Science, Kunming University of Science and Technology, Kunming, Yunnan 650500, China}

\author{Hoernisa Iminniyaz}
\email[Corresponding author:~]{wrns@xju.edu.cn}
\address{School of Physics Science and Technology, Xinjiang University, Urumqi 830046, China}

\title{Thermodynamics and phase transitions of spherically symmetric AdS regular black holes}

	\begin{abstract}
		We obtain the singular ``mother'' black hole solution by solving the coupling between Einstein gravity with a cosmological constant and a nonlinear electromagnetic field. The thermodynamic quantities of the ``mother'' black hole are first derived in the unconstrained phase space $(S, q, \alpha, P)$, after which the regularity condition $M = q^3/\alpha$ is imposed to obtain the thermodynamic quantities of a class of spherically symmetric AdS regular black holes self-consistently. We distinguish two categories of thermodynamic quantities: fundamental conjugate variables defined by the first law, and thermodynamic response functions constructed from them. For response functions involving second-order derivatives — in particular the heat capacity — we show that the order of constraint imposition and derivative calculation cannot be interchanged, and the correct procedure is to first impose the constraint and then construct the response function. Within this framework, we find that when $P<P_c$, the $G$--$T$ diagram exhibits a behavior qualitatively different from the swallowtail behavior of the standard RN-AdS black hole. Instead, it exhibits an 8-shaped structure for $P<P_z$ and a C-shaped structure for $P_z<P<P_c$, corresponding to first-order and zeroth-order phase transitions between small and large black hole phases, respectively. Imposing the regularity condition thus leads to a richer phase structure for this class of black holes than the standard RN-AdS case.

	\end{abstract}

	\maketitle

	\section{Introduction}
	\label{sec:1}

In 1970, Hawking and Penrose~\cite{Hawking:1970zqf} proved that any spacetime satisfying certain physically reasonable conditions must be causally geodesically incomplete, thereby implying the existence of a singularity, where curvature diverges and known physical laws break down. The singularity problem thus reveals the incompleteness of general relativity. Although a complete resolution of the singularity problem is generally expected to require a theory of quantum gravity, it is nevertheless important to explore whether singularities can be avoided within effective classical frameworks.

The first regular black hole solution was proposed by Bardeen in 1968~\cite{bardeen1968non}, which avoids singularities via a de Sitter core that generates negative pressure, thereby preventing a singular end-state of the gravitationally collapsed matter~\cite{Sakharov:1966aja}. A major advancement was achieved by Ay\'{o}n-Beato and Garc\'{i}a~\cite{Ayon-Beato:1998hmi,Ayon-Beato:2000mjt}, who constructed a regular black hole model within Einstein gravity coupled to a nonlinear electromagnetic field. They further showed that the Bardeen model can be interpreted within the same framework as the gravitational field corresponding to a nonlinear magnetic monopole. By considering different forms of the nonlinear electrodynamic Lagrangian, numerous regular black hole solutions were developed~\cite{Hayward:2005gi,Berej:2006cc,Burinskii:2002pz,Balart:2014cga,Ma:2015gpa,Nam:2018uvc}. In 2016, Fan et al.~\cite{Fan:2016hvf,Toshmatov:2018cks} obtained a class of regular black hole solutions by considering a general nonlinear electrodynamic Lagrangian, followed by the derivation of its rotating counterpart~\cite{Kar:2025anc}.

In parallel with geometric developments, the thermodynamics of black holes has grown into a rich field of study following the discovery of Bekenstein--Hawking entropy~\cite{Bekenstein:1973ur,Bekenstein:1974jk} and Hawking temperature~\cite{Hawking:1974rv,Hawking:1975vcx}. A particularly fruitful development in this direction is the extended phase space formulation, in which the cosmological constant is treated as a thermodynamic pressure
$P=-\Lambda/(8\pi)$, and the black hole mass is interpreted as enthalpy rather than internal energy~\cite{Creighton:1995au,Padmanabhan:2002sha}. Chamblin et al. first discovered that RN-AdS black holes undergo a 1-order phase transition analogous to the liquid/gas phase transition in a van der Waals fluid~\cite{Chamblin:1999tk,Chamblin:1999hg}. Kubiz\v{n}\'{a}k et al. further strengthened the relationship between RN-AdS black holes and van der Waals fluids through a detailed $P$--$V$ criticality analysis~\cite{Kubiznak:2012wp}. The resulting framework, now commonly referred to as black hole chemistry, has since been generalized to various other black hole spacetimes~\cite{Gunasekaran:2012dq,Abbas:2024pyi,Doneva:2010ntr,Kubiznak:2016qmn,Abdusattar:2023hlj,Kubiznak:2014zwa,Rafiee:2021hyj,Tzikas:2018cvs,Kumar:2020cve,Ali:2020bgc,Ali:2019rjn}.

Despite these advances, when the thermodynamic framework developed for standard black holes is applied to regular black holes, a fundamental inconsistency appears: the Hawking temperature computed from surface gravity is not equal to the temperature derived from the first law of thermodynamics~\cite{Ma:2014qma,Ma:2025dee}. This inconsistency arises from the regularity condition, which eliminates the central singularity, introducing a constraint among the black hole parameters. This constraint not only reduces the dimension of the thermodynamic phase space, but also modifies the first law of thermodynamics in that phase space. To address this issue, Ma et al.~\cite{Ma:2025dee,Ma:2025tzm,Xia:2026nsp} proposed a resolution and studied the phase transitions for the Bardeen-AdS and Hayward-AdS regular black holes. The key idea is to proceed in two steps: first, derive the full thermodynamic quantities for the singular ``mother'' black hole in the unconstrained phase space where all parameters are independent; second, impose the regularity condition on these quantities to obtain the thermodynamics of the regular black hole. This two-step procedure yields a self-consistent thermodynamic description where the Hawking temperature computed from surface gravity coincides with that derived from the first law.

Although the foregoing method has successfully resolved the temperature inconsistency, applying it to calculate the heat capacity and the Gibbs free energy in the same way (first deriving the unconstrained expressions for the ``mother'' black hole and then imposing the regularity condition) leads to a serious problem: the critical pressure $P_c$ obtained from the Gibbs free energy differs from that obtained from the heat capacity. For a self-consistent thermodynamic theory, this constitutes a fatal flaw.

We point out that this discrepancy arises because the heat capacity, being a second-order thermodynamic response function ($C_{P} \equiv dM/dT$), does not allow for the interchangeability of constraint imposition and differentiation. Computing the heat capacity by first differentiating the unconstrained mass $M_1$ with respect to the unconstrained temperature $T_1$ and then imposing the constraint yields $C_{P1}\bigr|_{\alpha=q^3/M}$, which differs from the correct quantity $C_{P2}=(dM_2/dT_2)_{q,P}$. The latter is obtained by first imposing the constraint $M=q^3/\alpha$ on $M_1$ and $T_1$, and then differentiating. By contrast, the Gibbs free energy is a first-order Legendre transform $G \equiv M - TS$, where the two orders give identical results.

In this paper, we generalize the foregoing method from specific models to a class of spherically symmetric AdS regular black holes: we construct the singular ``mother'' black hole solution from Einstein gravity coupled to nonlinear electrodynamics, derive the fundamental thermodynamic quantities of the ``mother'' black hole in the full unconstrained phase space $(S,q,\alpha,P)$, and impose the regularity condition $M=q^3/\alpha$ to obtain the fundamental thermodynamic quantities of a class of spherically symmetric AdS regular black holes. To eliminate the $P_c$ contradiction, we distinguish two categories of thermodynamic quantities: fundamental conjugate variables defined directly by the first law, for which the constraint is applied after differentiation; and thermodynamic response functions constructed from these conjugate variables together with the independent extensive variables, for which the constraint must be imposed on the fundamental quantities before constructing the response functions. This logical hierarchy is the key to ensuring the self-consistency of physical quantities such as the critical pressure.

Guided by the above considerations, we derive the heat capacity $C_{P2}$ and the Gibbs free energy $G_2$ using the constrained temperature $T_2$ and enthalpy $M_2$, and systematically investigate the phase structure of a class of spherically symmetric AdS regular black holes. Our results reveal that when $P<P_c$, the $G$--$T$ diagram exhibits two distinct structures separated by a characteristic pressure $P_z$: for $P<P_z$, an 8-shaped curve, and for $P_z<P<P_c$, a C-shaped curve, corresponding respectively to first-order and zeroth-order phase transitions between the small black hole (SBH) and large black hole (LBH) phases. The heat capacity $C_{P2}$ displays two divergences, marking two second-order phase transitions from the SBH to the intermediate black hole (IBH) and from the IBH to the LBH. The $P$--$V$ diagram confirms the liquid/gas-like character of the SBH--LBH transition, and analytic expressions for $T_c$ and $P_c$ are obtained.

The paper is organized as follows. In Sec.~\ref{sec:2}, we derive the singular ``mother'' black hole solution for a class of spherically symmetric AdS regular black holes. In Sec.~\ref{sec:3}, we obtain the fundamental thermodynamic quantities for both the ``mother'' and the regular black holes. In Sec.~\ref{sec:4}, we present the Gibbs free energy, heat capacity, and $P$--$V$ criticality for this class of black holes, and analyze their phase structure. In Sec.~\ref{sec:5}, we present conclusions and discussion.

	\section{The singular ``mother'' black hole}
	\label{sec:2}
In this section, we derive the singular ``mother'' black hole solution for a class of spherically symmetric AdS regular black holes. The coupling of Einstein gravity with a cosmological constant to a nonlinear electromagnetic field is described by the following field equations:
\begin{equation}\label{eq:21}
\begin{array}{c}
G_{\mu\nu} + \Lambda g_{\mu\nu} = T_{\mu\nu}, \qquad \nabla_\mu (\mathcal{L}_{\mathcal{F}} F^{\mu\nu}) = 0,
\end{array}
\end{equation}
where $\Lambda$ is the cosmological constant and $\mathcal{L}_{\mathcal{F}} \equiv \partial\mathcal{L}/\partial\mathcal{F}$. The nonlinear electromagnetic energy-momentum tensor takes the form
\begin{equation}\label{eq:22}
\begin{array}{c}
T_{\mu \nu} = 2\Bigl(\mathcal{L}_{\mathcal{F}} F_{\mu\rho}F_{\nu}^{\ \rho} - \tfrac{1}{4} g_{\mu\nu} \mathcal{L}\Bigr).
\end{array}
\end{equation}

The line element of a static spherically symmetric black hole is given by
\begin{equation}\label{eq:23}
\begin{array}{c}
ds^{2} = -f(r)dt^{2} + f(r)^{-1}dr^{2} + r^{2}(d\theta^{2} + \sin^{2}\theta\,d\phi^{2}),
\end{array}
\end{equation}
where the metric function is
\begin{equation}\label{eq:24}
\begin{array}{c}
f(r) = 1 - \frac{2m(r)}{r}.
\end{array}
\end{equation}
Substituting the above metric into Eq.~(\ref{eq:21}), we obtain the Lagrangian density as
\begin{equation}\label{eq:25}
\begin{array}{c}
\mathcal{L}(r) = \frac{4m'(r)}{r^{2}} - 2\Lambda.
\end{array}
\end{equation}
Integrating Eq.~(\ref{eq:25}) and imposing the boundary condition $\lim\limits_{r\to\infty}m(r)=M$, we obtain
\begin{equation}\label{eq:26}
\begin{array}{c}
m(r) = M - \frac{1}{4}\int_{r}^{\infty} r^{2} \mathcal{L}(r) \,dr+ \frac{\Lambda}{6}\,r^{3},
\end{array}
\end{equation}
where $M$ is the ADM mass in asymptotically flat spacetime.

To obtain the ``mother'' black hole solution for a class of AdS regular black holes, we adopt the nonlinear electrodynamics Lagrangian~\cite{Fan:2016hvf,Toshmatov:2018cks}:
\begin{equation}\label{eq:27}
\begin{array}{c}
\mathcal{L}(\mathcal{F}) = \frac{4\mu}{\alpha}\,
\frac{(\alpha\mathcal{F})^{\frac{\nu+3}{4}}}
{\bigl[1 + (\alpha\mathcal{F})^{\frac{\nu}{4}}\bigr]^{\frac{\mu+\nu}{\nu}}},
\end{array}
\end{equation}
where $\mathcal{F} \equiv \frac{1}{4}F_{\mu\nu}F^{\mu\nu} = 2Q_m^{2}/r^{4}$, with $Q_m$ the magnetic charge. The parameter $q$ is a regularization parameter related to $Q_m$ via $Q_m = q^{2}/\sqrt{2\alpha}$, and $\alpha>0$ has dimensions of length squared.

Substituting the Lagrangian density Eq.~(\ref{eq:27}) into Eq.~(\ref{eq:26}), we obtain
\begin{equation}\label{eq:28}
\begin{array}{c}
m(r) = M - \frac{q^{3}}{\alpha}\left(1 - \frac{r^{\mu}}{(r^{\nu} + q^{\nu})^{\mu/\nu}}\right) + \frac{\Lambda}{6}\,r^{3},
\end{array}
\end{equation}
where $\mu\ge 3$. Substituting Eq.~(\ref{eq:28}) into Eq.~(\ref{eq:24}), the metric function of the singular ``mother'' black hole is
\begin{equation}\label{eq:210}
\begin{array}{c}
f(r) = 1 - \frac{2}{r}\left(M - \frac{q^{3}}{\alpha}\left(1 - \frac{r^{\mu}}{(r^{\nu} + q^{\nu})^{\mu/\nu}}\right)\right) - \frac{\Lambda}{3}\,r^{2}.
\end{array}
\end{equation}
When $\Lambda=0$, the above equation reduces to the asymptotically flat ``mother'' solution discussed in Refs.~\cite{Fan:2016hvf,Li:2025constraint}.

\begin{figure*}
\centering
\begin{tabular}{c c}
\includegraphics[scale=0.55]{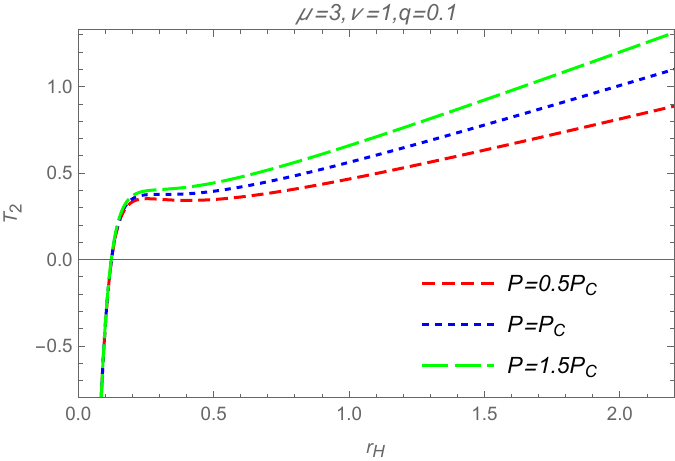}\hspace{-0.05cm}
&\includegraphics[scale=0.55]{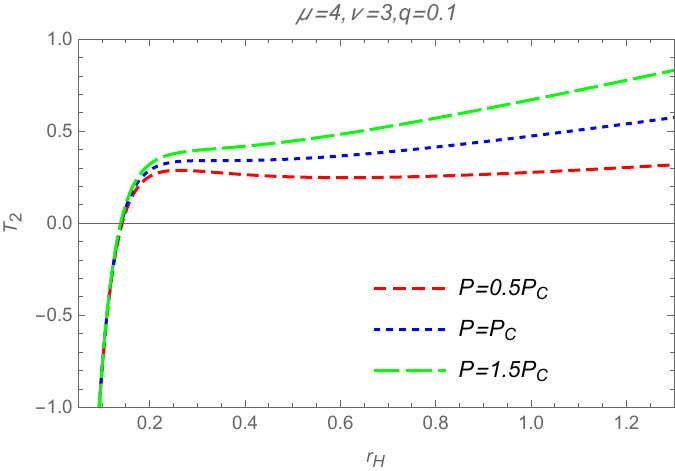}
\end{tabular}
\caption{The behavior of $T_2$ vs $r_H$ for different parameters $\mu$, $\nu$ and $P$ with $q=0.1$.}
\label{fig:T-r}
\end{figure*}
The regularity condition is $\lim\limits_{r \to 0} m(r) = 0$~\cite{Fan:2016hvf,Ma:2025dee}. Substituting Eq.~(\ref{eq:28}) into this condition, we obtain
\begin{equation}\label{eq:211}
\begin{array}{c}
M = \frac{q^{3}}{\alpha},
\end{array}
\end{equation}
where the cosmological constant $\Lambda$ does not affect this regularity condition.

Imposing the regularity condition $M = \frac{q^{3}}{\alpha}$ on the ``mother'' black hole metric function Eq.~(\ref{eq:210}), the solution reduces to a class of spherically symmetric AdS regular black holes:
\begin{equation}\label{eq:212}
\begin{array}{c}
f(r) = 1 - \frac{2M\,r^{\mu-1}}{(r^{\nu} + q^{\nu})^{\mu/\nu}} - \frac{\Lambda}{3}\,r^{2}.
\end{array}
\end{equation}
For Eq.~(\ref{eq:212}), when $\mu=3$ and $\nu=1$, it reduces to the Fan-Wang-AdS regular black hole~\cite{Fan:2016hvf}, when $\mu=3$ and $\nu=2$, it reduces to the Bardeen-AdS black hole, and when $\mu=\nu=3$, it reduces to the Hayward-AdS black hole.

	\section{Thermodynamics of Regular AdS Black Holes}
	\label{sec:3}
In this section, we first derive the thermodynamic quantities of the ``mother'' black hole in the full unconstrained phase space, and then impose the regularity condition $M = q^3/\alpha$ on these quantities to obtain the self-consistent thermodynamic quantities for a class of spherically symmetric AdS regular black holes.

\subsection{Thermodynamics of the singular ``mother'' black hole}

Using the metric function Eq.~(\ref{eq:210}) and the event horizon condition $f(r_H)=0$, the black hole mass can be expressed in terms of the event horizon radius $r_H$ as
\begin{equation}\label{eq:31}
\begin{array}{c}
M = \frac{r_H}{2}\left(1 - \frac{\Lambda}{3} r_H^{2}\right) + \frac{q^3}{\alpha}\left(1 - \frac{r_H^{\mu}}{(r_H^{\nu} + q^{\nu})^{\mu/\nu}}\right).
\end{array}
\end{equation}
In the extended phase space, the cosmological constant is interpreted as the thermodynamic pressure $P = -\Lambda/(8\pi)$~\cite{Creighton:1995au,Padmanabhan:2002sha}. According to the Bekenstein--Hawking area law $S = \pi r_H^{2}$, the mass can be expressed in the full phase space $(S, q, \alpha, P)$:
\begin{equation}\label{eq:32}
\begin{array}{c}
M = \frac{1}{2}\sqrt{\frac{S}{\pi}}\left(1 + \frac{8}{3}PS\right) + \frac{q^3}{\alpha}\left(1 - \frac{\left(\frac{S}{\pi}\right)^{\mu/2}}{\left(\left(\frac{S}{\pi}\right)^{\nu/2} + q^{\nu}\right)^{\mu/\nu}}\right).
\end{array}
\end{equation}
In the unconstrained phase space, all four parameters $(S, q, \alpha, P)$ are treated as independent thermodynamic variables. The first law of thermodynamics reads
\begin{equation}\label{eq:33}
\begin{array}{c}
dM = T_1 dS + \Phi_1 dq + \mathcal{A}_1 d\alpha + V dP,
\end{array}
\end{equation}
where the conjugate thermodynamic quantities are
\begin{equation}\label{eq:34}
\begin{aligned}
T_1 &\equiv \left(\frac{\partial M}{\partial S}\right)_{q,\alpha,P}, &
\Phi_1 &\equiv \left(\frac{\partial M}{\partial q}\right)_{S,\alpha,P},\\
\mathcal{A}_1 &\equiv \left(\frac{\partial M}{\partial \alpha}\right)_{S,q,P}, &
V &\equiv \left(\frac{\partial M}{\partial P}\right)_{S,q,\alpha},
\end{aligned}
\end{equation}
where the subscript ``1'' indicates that these quantities refer to the singular ``mother'' black hole. The thermodynamic volume $V$ carries no subscript because $V$ does not involve the parameters $\alpha$ and $q$ in the regularity condition $M = q^3/\alpha$, and is therefore the same for both the ``mother'' and the regular black hole. From Eqs.~(\ref{eq:32}) and (\ref{eq:34}), the thermodynamic quantities of the singular ``mother'' black hole are obtained as:
\begin{equation}\label{eq:35}
\begin{aligned}
T_1 &= \dfrac{1}{4\pi r_H} + 2P r_H - \dfrac{q^{3}}{2\pi\alpha\,r_H^{2}}\,A^{-1}B,\\
\Phi_1 &= \dfrac{q^2}{\alpha}\Bigl(3 + A^{-1}(-3 + B)\Bigr),\\
\mathcal{A}_1 &= -\dfrac{q^3}{\alpha^2}\bigl(1 - A^{-1}\bigr), \qquad
V = \dfrac{4}{3}\pi r_H^{3},
\end{aligned}
\end{equation}
where $A = (1 + q^{\nu}/r_H^{\nu})^{\mu/\nu}$ and $B = \mu q^{\nu}/(r_H^{\nu} + q^{\nu})$. It can be verified that the thermodynamic quantities in Eq.~(\ref{eq:35}) together with the mass function Eq.~(\ref{eq:32}) satisfy the Smarr relation
\begin{equation}\label{eq:38}
\begin{aligned}
M = 2T_1 S - 2V P + \Phi_1 q + 2\mathcal{A}_1 \alpha.
\end{aligned}
\end{equation}
\subsection{Thermodynamics of the AdS regular black hole}
\begin{figure*}
\centering
\begin{tabular}{c c}
\includegraphics[scale=0.55]{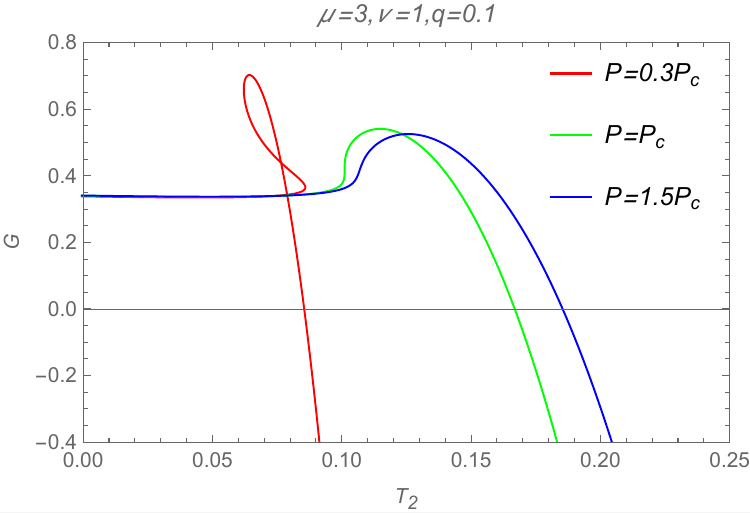}\hspace{-0.2cm}
&\includegraphics[scale=0.55]{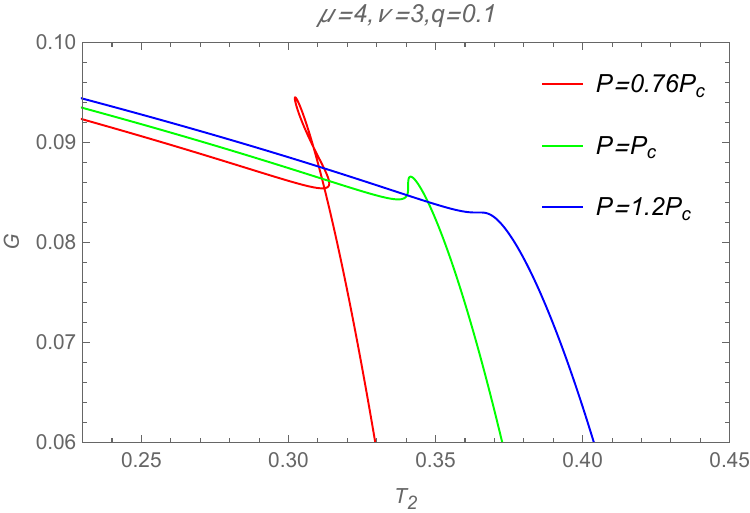}
\end{tabular}
\caption{The behavior of $G_2$ vs $T_2$ for different $\mu$, $\nu$ and $P$ with $q=0.1$.}
\label{fig:G-T}
\end{figure*}

We now impose the regularity condition $M = q^3/\alpha$ on the thermodynamic quantities of the ``mother'' black hole, thereby deriving the thermodynamic quantities of a class of AdS regular black holes. Substituting the regularity condition into Eq.~(\ref{eq:35}), we obtain
\begin{equation}\label{eq:39}
\begin{array}{c}
T_2 = \frac{3 \left(8 \pi  P r_H^2+1\right) r_H^{\nu }+q^{\nu } \left(-3 \mu -8 \pi  (\mu -3) P r_H^2+3\right)}{12 \pi  r_H \left(q^{\nu }+r_H^{\nu }\right)},
\end{array}
\end{equation}
\begin{equation}\label{eq:310}
\begin{array}{c}
\Phi_2 = \frac{r_H\left(3 + 8\pi P r_H^{2}\right)}{6q}\left(3(A - 1) + B\right),
\end{array}
\end{equation}
\begin{equation}\label{eq:311}
\begin{array}{c}
\mathcal{A}_2 =- \frac{r_H^{2}\left(3 + 8\pi P r_H^{2}\right)^{2}}{36q^{3}}\left(A^{2} - A\right),
\end{array}
\end{equation}
where the subscript ``2'' indicates that these quantities refer to the regular black hole. Previously, we have obtained various thermodynamic quantities for a class of AdS regular black holes. Now we specifically investigate the properties of the Hawking temperature $T_2$. The behavior of the Hawking temperature $T_2$ with respect to the event horizon $r_H$ for various values of the pressure $P$ is shown in Fig.~\ref{fig:T-r}, where the left and right panels correspond to the ($\mu=3$, $\nu=1$) and ($\mu=4$, $\nu=3$) cases, respectively. When the local maximum and minimum values of the Hawking temperature coincide, the Hawking temperature satisfies the following conditions:
\begin{equation}\label{eq:Tcrit}
\begin{aligned}
\frac{\partial T_2}{\partial r_H}=0 \quad \text{and} \quad \frac{\partial^2 T_2}{\partial r_H^2}=0.
\end{aligned}
\end{equation}
From Eq.~(\ref{eq:Tcrit}), for given values of $q$, $\mu$, and $\nu$, the critical pressure $P_c$ can be obtained numerically. By analyzing Eq.~(\ref{eq:39}) and Fig.~\ref{fig:T-r}, we get the following conclusions:
\begin{itemize}
\item When $P<P_c$, the Hawking temperature has a local maximum at $r_H=r_{H1}$ and a local minimum at $r_H=r_{H2}$, thereby partitioning the black hole into three branches: the SBH branch for $0<r_H<r_{H1}$, the IBH branch for $r_{H1}<r_H<r_{H2}$, and the LBH branch for $r_H>r_{H2}$. For example, for the Fan-Wang-AdS black hole with $q=0.1$ and $P=0.0122553<P_c=0.0245106$ (the red dashed curve in Fig.~\ref{fig:T-r}), the local maximum and minimum occur at $r_{H1}=0.492306$ and $r_{H2}=1.41979$, respectively. Correspondingly, in the SBH branch ($0<r_H<0.492306$) and the LBH branch ($r_H>1.41979$), the Hawking temperature increases as $r_H$ increases, whereas in the IBH branch ($0.492306<r_H<1.41979$), it decreases as $r_H$ increases.
\item When $P=P_c$, the local maximum and minimum coalesce into a single inflection point (the blue dashed curve in Fig.~\ref{fig:T-r}). For $P>P_c$, the Hawking temperature increases monotonically as $r_H$ increases (the green dashed curve in Fig.~\ref{fig:T-r}).
\end{itemize}
The properties of the Hawking temperature $T_2$ described above are closely related to the phase structure and stability of the black hole, and its local structure governs the phase behavior of the system.

\begin{figure*}
\centering
\subfloat[][]{\includegraphics[scale=0.465]{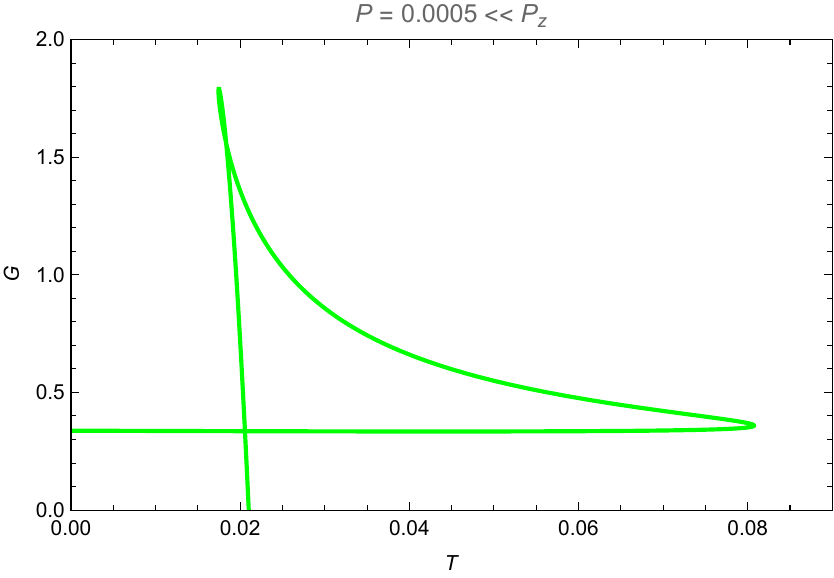}}\hspace{0cm}
\subfloat[][]{\includegraphics[scale=0.47]{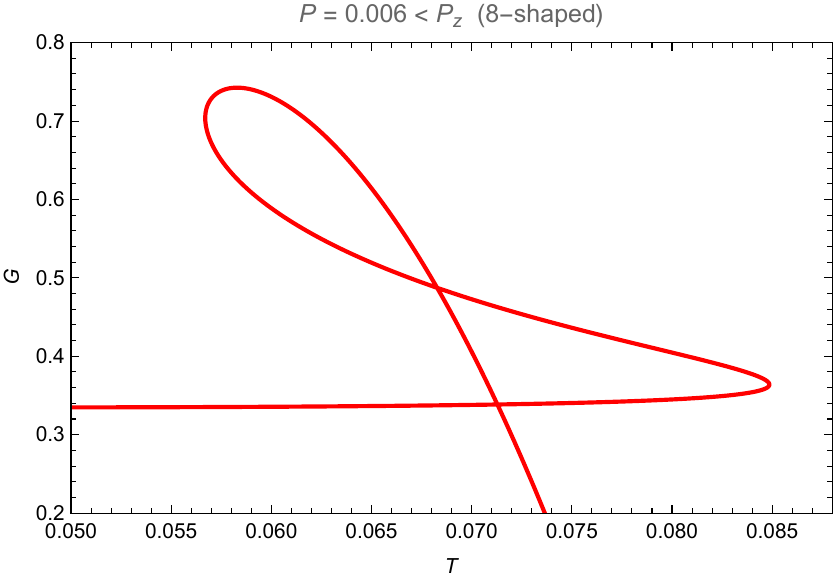}}\\[2mm]
\subfloat[][]{\includegraphics[scale=0.46]{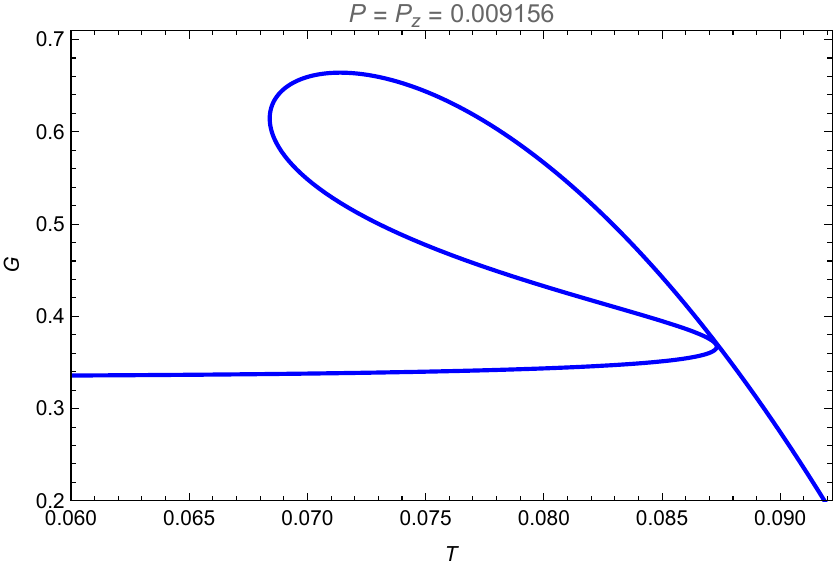}}\hspace{0cm}
\subfloat[][]{\includegraphics[scale=0.472]{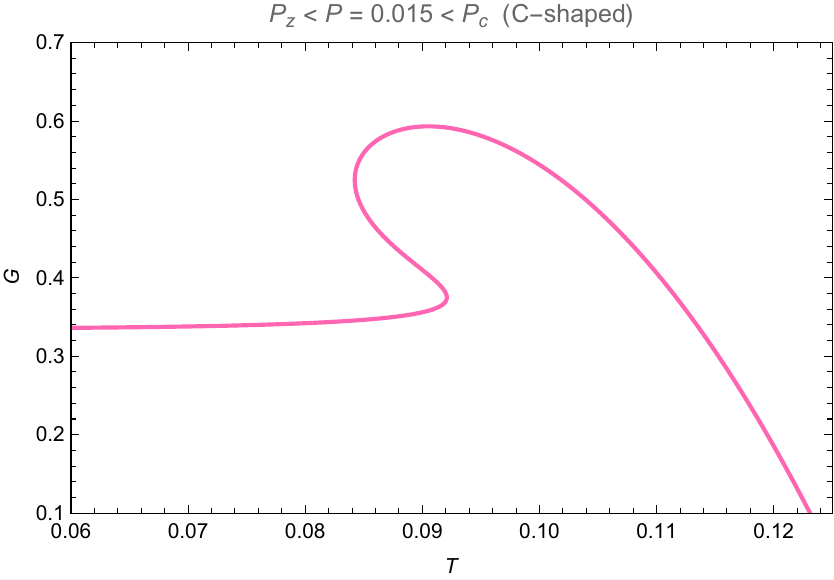}}
\caption{The behavior of $G_2$ vs $T_2$ for the Fan-Wang-AdS black hole ($\mu=3$, $\nu=1$) with $q=0.1$, at various $P$.}
\label{fig:G-T4}
\end{figure*}

\section{Phase transitions and thermodynamic stability}
\label{sec:4}
In this section, we investigate the thermodynamic stability and phase structure of a class of AdS regular black holes. Under the constraint $M = q^3/\alpha$, thermodynamic quantities must be handled in a specific logical order. They fall into two conceptually distinct categories.

The first category comprises the fundamental conjugate quantities defined directly by the first law — the Hawking temperature $T$, the electric potential $\Phi$, the thermodynamic quantity $\mathcal{A}$, and the thermodynamic volume $V$ — which are obtained as partial derivatives of the unconstrained mass function $M(S,q,\alpha,P)$. For the regular black hole, these quantities are obtained by first taking the derivatives and then imposing the constraint $M = q^3/\alpha$, thereby yielding $T_2$, $\Phi_2$, $\mathcal{A}_2$, and $V$.

The second category comprises thermodynamic response functions, which are constructed from the fundamental conjugate quantities of the first category together with the independent extensive parameters $(S,q,\alpha,P)$. For response functions involving first-order constructions — such as the Gibbs free energy $G = M - TS$ — the operations of constraint imposition and Legendre transformation are interchangeable: whether the constraint is imposed before or after the transformation, the result is the same. For response functions involving second-order derivatives — in particular the heat capacity $C_{P} \equiv (dM/dT)_{q,P}$ — the two operations are not interchangeable: one must first impose the constraint on the fundamental quantities and then perform the differentiation. Reversing the order yields an incorrect result. Therefore, the thermodynamic response functions in the constrained phase space must be constructed directly from the constrained fundamental thermodynamic quantities, rather than obtained by imposing the constraint on the response functions computed in the higher-dimensional unconstrained phase space.
\subsection{Gibbs free energy and phase transitions}
\label{sec:4.1}
The Gibbs free energy is the thermodynamic potential that governs phase transitions at fixed pressure. As emphasized above, $G$ is a thermodynamic response function, defined by the Legendre transformation $G = M - TS$. For a class of spherically symmetric AdS regular black holes, it must be constructed from the constrained mass $M_2$ and temperature $T_2$. Substituting $M = q^3/\alpha$ into Eq.~(\ref{eq:32}) yields the constrained mass:
\begin{equation}
\begin{array}{c}
M_2 = \dfrac{1}{2}\sqrt{\dfrac{S}{\pi}}\Bigl(1 + \dfrac{8}{3} P S\Bigr)\Bigl(1 + \dfrac{q^{\nu}}{(S/\pi)^{\nu/2}}\Bigr)^{\mu/\nu}.
\end{array}
\label{eq:40}
\end{equation}
Inserting Eq.~(\ref{eq:40}) and $T_2$ into the definition of the Gibbs free energy, we obtain
\begin{equation}\label{eq:41}
\begin{aligned}
G_2 &= M_2 - T_2 S \\
&= \dfrac{1}{6} r_H^{1-\mu}\bigl(3+8\pi P r_H^{2}\bigr)\bigl(q^{\nu}+r_H^{\nu}\bigr)^{\mu/\nu} \nonumber\\
&\quad - \dfrac{r_H\Bigl[3r_H^{\nu}\bigl(1+8\pi P r_H^{2}\bigr)+q^{\nu}\bigl(3-3\mu-8\pi(\mu-3)P r_H^{2}\bigr)\Bigr]}{12\bigl(q^{\nu}+r_H^{\nu}\bigr)},
\end{aligned}
\end{equation}
which is the Gibbs free energy of a class of AdS regular black holes. The phase structure is analyzed through the behavior of $G_2$ with respect to $T_2$ for different parameters $\mu$, $\nu$ and $P$ with $q=0.1$, as shown in Fig.~\ref{fig:G-T}, where the left and right panels correspond to the ($\mu=3$, $\nu=1$) and ($\mu=4$, $\nu=3$) cases respectively. When $P<P_c$, the $G_2$--$T_2$ curves deviate qualitatively from the standard swallowtail structure of the RN-AdS black hole~\cite{Kubiznak:2012wp}. At $P=P_c$, the curves develop an inflection point, marking a second-order critical point. For $P>P_c$, no phase transition occurs.

\begin{figure*}
\centering
\begin{tabular}{c c}
\includegraphics[scale=0.5]{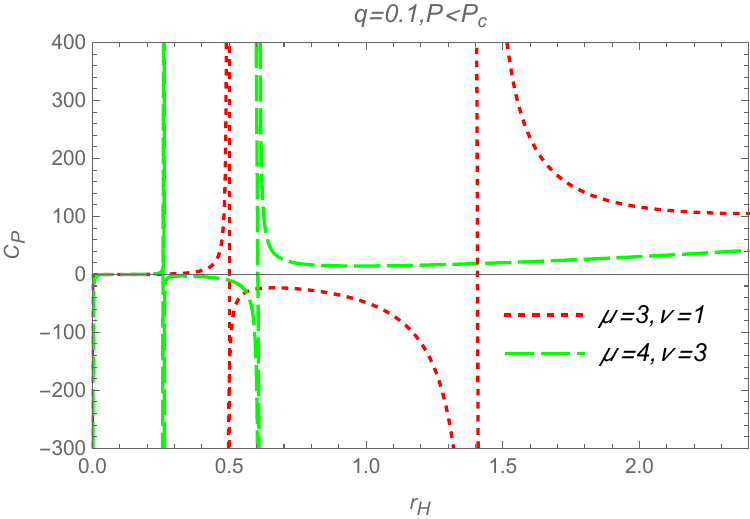}\hspace{-0.1cm}
&\includegraphics[scale=0.53]{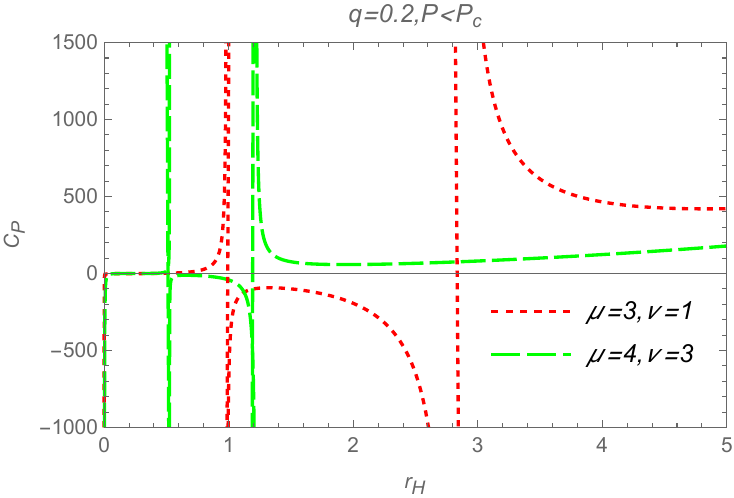}
\end{tabular}
\caption{The behavior of $C_{P2}$ vs $r_H$ for different $\mu$, $\nu$ and $q$ at $P<P_c$.}
\label{fig:C-P1}
\end{figure*}

\begin{figure*}
\centering
\begin{tabular}{c c}
\includegraphics[scale=0.5]{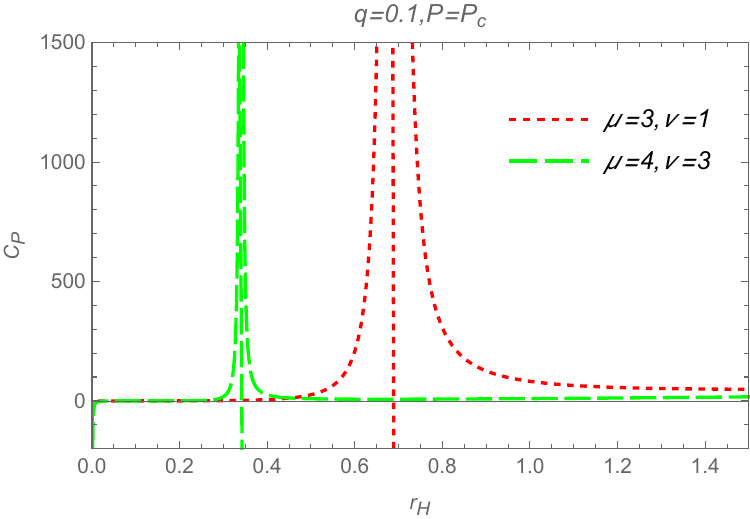}\hspace{-0.1cm}
&\includegraphics[scale=0.5]{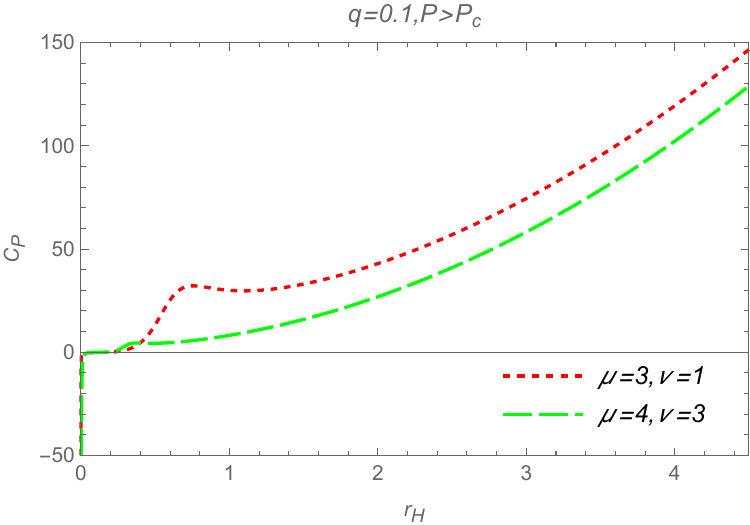}
\end{tabular}
\caption{The behavior of $C_{P2}$ vs $r_H$ for $P=P_c$ (left) and $P>P_c$ (right), with $q=0.1$.}
\label{fig:C-P2}
\end{figure*}

To examine the subcritical behavior in greater detail, we focus on the ($\mu=3$, $\nu=1$) case with $q=0.1$, for which $P_c=0.0245106$. In addition to $P_c$, there exists a characteristic pressure $P_z=0.009156<P_c$. The $G_2$--$T_2$ curves at four representative pressures are shown in Fig.~\ref{fig:G-T4}. When $P<P_z$, the curve closes into an 8-shaped structure whose self-intersection point marks a first-order SBH--LBH phase transition (Fig.~\ref{fig:G-T4}b). When $P\ll P_z$, the upper loop is small and the lower loop is large, giving the 8-shape a swallowtail-like appearance (Fig.~\ref{fig:G-T4}a). As $P$ increases towards $P_z$, the upper loop enlarges while the lower loop shrinks. At $P=P_z$, the lower loop disappears and the curve becomes a single-loop zero-like shape (Fig.~\ref{fig:G-T4}c). When $P_z<P<P_c$, the curve becomes a C-shaped structure (Fig.~\ref{fig:G-T4}d). In this regime, the Gibbs free energy undergoes a discontinuous jump, signaling a zeroth-order phase transition. The above analysis exhibits that this class of regular black holes exhibits a richer phase structure involving both first-order and zeroth-order phase transitions, which originates from the constraint-induced modifications in the thermodynamic phase space~\cite{Ma:2025dee}.

\subsection{Heat capacity and local thermodynamic stability}
\label{sec:4.2}
\begin{figure*}
\centering
\begin{tabular}{c c}
\includegraphics[scale=0.5]{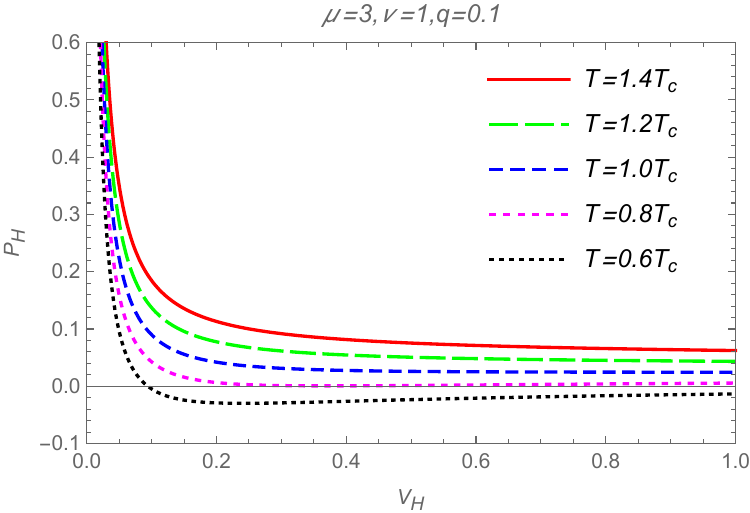}\hspace{-0.1cm}
&\includegraphics[scale=0.51]{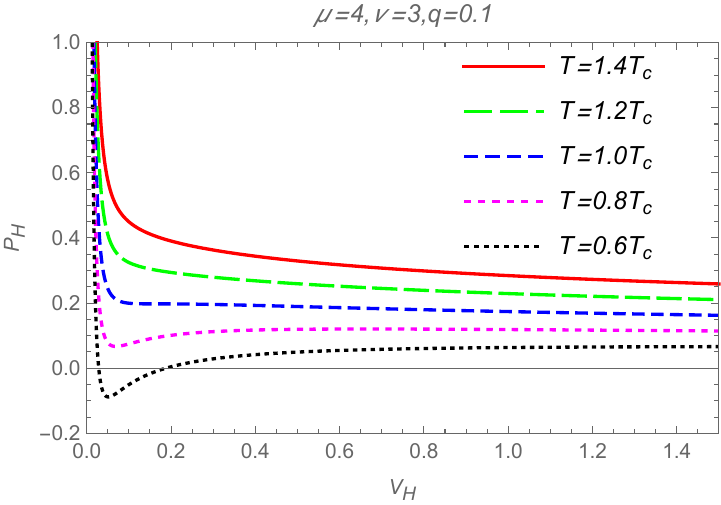}
\end{tabular}
\caption{The behavior of $P$ vs $V$ for different $\mu$, $\nu$ and $T$ with $q=0.1$.}
\label{fig:P-V}
\end{figure*}
The local thermodynamic stability of a black hole can be examined via its heat capacity at fixed pressure and charge,
\begin{equation}\label{eq:43}
C_P \equiv \left(\dfrac{dM}{dT}\right)_{P,\,q}.
\end{equation}
A branch with $C_P>0$ is locally thermodynamically stable, whereas $C_P<0$ indicates local thermodynamic instability.

For the regular black holes, the heat capacity must be evaluated using the constrained thermodynamic quantities $M_2$ and $T_2$. Equivalently, at fixed $P$ and $q$, $M_2$ and $T_2$ can be treated as functions of the horizon radius $r_H$, so that
\begin{equation}\label{eq:Cchain}
C_{P2}
= \left(\dfrac{dM_2}{dT_2}\right)_{q,P}
= \frac{dM_2/dr_H}{dT_2/dr_H}.
\end{equation}
This expression makes clear that the divergences of $C_{P2}$ occur at the stationary points of the Hawking temperature, namely at the points satisfying $dT_2/dr_H=0$, provided that $dM_2/dr_H$ remains finite and nonzero.

Substituting the constrained mass $M_2$ and the constrained temperature $T_2$ into Eq.~(\ref{eq:Cchain}), we obtain
\begin{equation}\label{eq:44}
\begin{aligned}
 C_{P2} &= \left(\dfrac{dM_2}{dT_2}\right)_{q,P} \\
 &= 2\pi r_H^{2-\mu}\bigl(q^{\nu}+r_H^{\nu}\bigr)^{\frac{\mu+\nu}{\nu}}
   \times \Bigl(q^{\nu}\bigl(8\pi P r_H^{2}(\mu-3) \\
 &\quad +3(\mu-1)\bigr)
   -3r_H^{\nu}(1+8\pi P r_H^{2})\Bigr)\\
 &
   \times \Bigl(q^{2\nu}\bigl(3+8\pi P r_H^{2}(\mu-3)-3\mu\bigr) -3r_H^{2\nu}(8\pi P r_H^{2}-1)\\
 &
   - q^{\nu}r_H^{\nu}\bigl(8\pi P r_H^{2}(6+\mu(\nu-1))+3(\mu+\mu\nu-2)\bigr)\Bigr)^{-1}.
\end{aligned}
\end{equation}
In particular, $C_{P2}$ is not equal to the result $C_{P1}|_{\alpha = q^3/M}$ obtained by first computing $C_{P1}$ from the unconstrained derivatives and then imposing the constraint, reflecting the non-interchangeability of constraint imposition and derivative calculation for the heat capacity.

The divergence structure of $C_{P2}$ is directly controlled by the shape of the $T_2$--$r_H$ curve. As discussed in the previous section regarding the Hawking temperature $T_2$, when $P<P_c$, $T_2$ has two local extrema: a local maximum at smaller $r_H$ and a local minimum at larger $r_H$. Since $C_{P2}=(dM_2/dr_H)/(dT_2/dr_H)$, these two local extrema correspond to two divergences of the heat capacity, separating the small black hole (SBH), intermediate black hole (IBH), and large black hole (LBH) branches. This behavior is confirmed by Fig.~\ref{fig:C-P1}, where the left and right panels correspond to $q=0.1$ and $q=0.2$, respectively. In each panel, the red and green curves represent the Fan-Wang-AdS ($\mu=3,\nu=1$) and ($\mu=4,\nu=3$) cases. For example, in the Fan-Wang-AdS black hole with $q=0.1$, the critical pressure is $P_c=0.0245106$. When $P=0.5P_c<P_c$, the heat capacity diverges at $r_{H1}=0.492306$ and $r_{H2}=1.41979$, which correspond, respectively, to the local maximum and the local minimum of the Hawking temperature $T_2$. These two divergences mark two second-order phase transitions: the SBH--IBH transition at $r_{H1}$ and the IBH--LBH transition at $r_{H2}$. The heat capacity is positive for $r_H<r_{H1}$ and $r_H>r_{H2}$, indicating that the SBH and LBH branches are locally stable, whereas $C_{P2}<0$ for $r_{H1}<r_H<r_{H2}$, showing that the IBH branch is locally unstable.

The behavior for $P\geq P_c$ is illustrated in Fig.~\ref{fig:C-P2}. The left panel corresponds to $P=P_c$, where the two local extrema of $T_2$ coalesce into a single inflection point and the two divergences of $C_{P2}$ merge at $r_H=0.687298$, marking a second-order phase transition. Here $C_{P2}>0$ and the black hole is stable. The right panel corresponds to $P>P_c$, where $T_2$ increases monotonically and $C_{P2}$ exhibits no divergence. In all regimes — whether $P<P_c$, $P=P_c$, or $P>P_c$ — the heat capacity remains negative near the origin, indicating that the very small black hole is always locally thermodynamically unstable.

\subsection{$P$--$V$ criticality}

From Eq.~(\ref{eq:39}), we can obtain the relationship between pressure and Hawking temperature as follows:
\begin{equation}\label{eq:45}
\begin{array}{c}
P = -\dfrac{3\bigl(-r_H^{\nu} - q^{\nu} + 4\pi q^{\nu} r_H T_2 + 4\pi r_H^{1+\nu} T_2 + q^{\nu}\mu\bigr)}
{8\pi r_H^{2}\bigl(-3r_H^{\nu} - 3q^{\nu} + q^{\nu}\mu\bigr)}.
\end{array}
\end{equation}
In the $P$--$V$ diagram, when the local maximum and minimum of the isotherm coalesce into a single inflection point, the system is at the critical point. This point satisfies
\begin{equation}\label{eq:46}
\begin{array}{c}
\left(\dfrac{\partial P}{\partial r_H}\right)_{T_2,q} = 0,
\qquad
\left(\dfrac{\partial^{2} P}{\partial r_H^{2}}\right)_{T_2,q} = 0.
\end{array}
\end{equation}
Solving Eqs.~(\ref{eq:45}) and (\ref{eq:46}), we obtain the critical temperature and critical pressure:
\begin{equation}\label{eq:47}
\begin{array}{c}
T_c = \dfrac{3r_H^{2\nu} + q^{2\nu}(3 - 4\mu + \mu^{2}) + q^{\nu}r_H^{\nu}\bigl(6 - \mu(4+\nu)\bigr)}
{2\pi r_H\bigl(3r_H^{2\nu} - q^{2\nu}(\mu-3) + q^{\nu}r_H^{\nu}(6 + \mu(\nu-1))\bigr)},
\end{array}
\end{equation}
\begin{equation}\label{eq:48}
\begin{array}{c}
P_c = \dfrac{3\bigl(-r_H^{2\nu} + q^{2\nu}(\mu-1) + q^{\nu}r_H^{\nu}(\mu + \mu\nu - 2)\bigr)}
{8\pi r_H^{2}\bigl(-3r_H^{2\nu} + q^{2\nu}(\mu-3) - q^{\nu}r_H^{\nu}(6 + \mu(\nu-1))\bigr)}.
\end{array}
\end{equation}
For the Fan-Wang-AdS ($\mu=3$, $\nu=1$) and ($\mu=4$, $\nu=3$) black holes, the critical values can be obtained by solving these equations numerically. Taking $q=0.1$ as an example, we obtain $T_c=0.101077$ and $P_c=0.0245106$ for the Fan-Wang-AdS black hole, and $T_c=0.340690$ and $P_c=0.197515$ for the ($\mu=4$, $\nu=3$) black hole.

The behavior of $P$ with respect to $V$ for various values of the Hawking temperature $T_2$ is shown in Fig.~\ref{fig:P-V}, where the left and right panels correspond to the Fan-Wang-AdS ($\mu=3$, $\nu=1$) and ($\mu=4$, $\nu=3$) cases, respectively. When $T<T_c$ (the pink and black dashed curves), the isotherms contain an oscillatory segment with a local minimum and a local maximum, indicating an SBH--LBH phase transition analogous to the liquid/gas phase transition in a van der Waals fluid. The physical isotherm is determined by the Maxwell equal-area construction, under which the SBH and LBH phases coexist. When $T=T_c$ (the blue dashed curve), the two extrema coalesce into an inflection point, marking the second-order critical point. When $T>T_c$ (the red and green dashed curves), the isotherms behave as an ideal gas.

\section{Conclusions and discussion}
	\label{sec:5}

In this paper, we have systematically investigated the thermodynamics and phase transitions of a class of spherically symmetric AdS regular black holes. By carefully distinguishing the logical roles of fundamental thermodynamic quantities and response functions, we have established a self-consistent thermodynamic description for a class of spherically symmetric AdS regular black holes. Our main results can be summarized as follows.

We have first obtained the singular ``mother'' black hole solution by solving the field equations of Einstein gravity with a negative cosmological constant coupled to nonlinear electrodynamics. The parameters $M$, $q$, and $\alpha$ in the metric function are independent. When the regularity condition $M = q^3/\alpha$ is imposed on the metric function, the ``mother'' black hole reduces to a class of spherically symmetric AdS regular black holes.

In the thermodynamic analysis, we have first obtained the thermodynamic quantities of the ``mother'' black hole from the first law in the full unconstrained phase space $(S, q, \alpha, P)$, and have then imposed the constraint $M = q^3/\alpha$ on these quantities to obtain the physical thermodynamic quantities $T_2$, $\Phi_2$, $\mathcal{A}_2$, and $V$ of the regular black hole. In particular, the Hawking temperature $T_2$ exhibits a local maximum and a local minimum below the critical pressure $P_c$, a behavior that underlies the phase structure found in this work.

A central methodological point of this work is the distinction between two categories of thermodynamic quantities. The first category consists of the conjugate quantities defined by the first law, for which the constraint is applied after differentiation. The second category comprises thermodynamic response functions constructed from those conjugate quantities together with the independent extensive parameters. For response functions involving first-order constructions — such as the Gibbs free energy $G = M - TS$ — the operations of constraint imposition and Legendre transformation are interchangeable: whether the constraint is imposed before or after the transform, the result is the same. For response functions involving second-order derivatives — in particular the heat capacity $C_P \equiv (dM/dT)_{q,P}$ — the two operations are not interchangeable: one must first impose the constraint on the fundamental quantities and then differentiate. Reversing the order yields an incorrect result. Therefore, the thermodynamic response functions in the constrained phase space must be constructed directly from the constrained fundamental thermodynamic quantities, rather than obtained by imposing the constraint on the response functions computed in the higher-dimensional unconstrained phase space.

The phase structure has been explored through the Gibbs free energy, the heat capacity, and the $P$--$V$ diagram. The overall behavior is governed by the critical pressure $P_c$. For $P<P_c$, the system exhibits three branches — SBH, IBH, and LBH — and the $P$--$V$ isotherms display oscillatory behavior characteristic of a liquid/gas-like phase transition. Moreover, there exists a characteristic pressure $P_z<P_c$ that divides the subcritical regime into two qualitatively distinct regions. For $P<P_z$, the $G_2$--$T_2$ curve forms an 8-shaped structure, signaling a first-order SBH--LBH phase transition. For $P_z<P<P_c$, the curve takes a C-shaped form, corresponding to a zeroth-order SBH--LBH phase transition in which the Gibbs free energy undergoes a discontinuous jump. At $P=P_c$, the system reaches a second-order critical point, and for $P>P_c$, no SBH--LBH phase transition occurs.

The local thermodynamic stability is determined by the heat capacity $C_{P2}$. For $P<P_c$, the Hawking temperature $T_2$ has a local maximum and a local minimum, and $C_{P2}$ correspondingly develops two divergences. These two divergences separate the SBH, IBH, and LBH branches. The SBH and LBH branches are locally stable with $C_{P2}>0$, whereas the IBH branch is locally unstable with $C_{P2}<0$. At $P=P_c$, the two extrema of $T_2$ coalesce and the two divergences of $C_{P2}$ merge. For $P>P_c$, $T_2$ becomes monotonic and no divergence appears in $C_{P2}$. Near the origin, $C_{P2}$ remains negative in all cases, indicating that very small regular black holes are always locally thermodynamically unstable.

Overall, the regularity condition not only eliminates the central singularity, but also reshapes the thermodynamic phase space. As a result, a class of spherically symmetric AdS regular black holes exhibits a richer phase structure than the standard RN-AdS case, including 8-shaped and C-shaped Gibbs free energy curves, first-order and zeroth-order SBH--LBH transitions, and constraint-induced modifications of local thermodynamic stability.

	\subsection*{Acknowledgments}
	The work is supported by the National Natural Science Foundation of China (Grant No. 12463001), and the Yunnan Xingdian Talent Support Program - Young Talent Project.


\end{document}